\documentclass[preprint2]{aastex}
\usepackage{graphicx}

\def\ltsima{$\; \buildrel < \over \sim \;$}
\def\simlt{\lower.5ex\hbox{\ltsima}}
\def\gtsima{$\; \buildrel > \over \sim \;$}
\def\simgt{\lower.5ex\hbox{\gtsima}}
\def\kms{{\rm\,km\,s^{-1}}}

\def\kpc{{\rm\,kpc}}

\def\msun{{\rm\,M_\odot}}

\def\pc{{\rm\,pc}}

\def\yr{{\rm\,yr}}

\newcommand{\fmmm}[1]{\mbox{$#1$}}

\newcommand{\scnp}{\mbox{\fmmm{''}}}
\newcommand{\mcnd}{\mbox{\fmmm{'}\hskip-0.3em .}}
\def\Aa{\; \buildrel \circ \over {\rm A}}
\def\AA{$\; \buildrel \circ \over {\rm A}$}
\def\yr{{\rm yr}}

\def\s{\ifmmode \widetilde \else \~\fi}
\def\={\overline}

\def\spose#1{\hbox to 0pt{#1\hss}}

\def\etal{{\it et al.\ }}

\def\lta{\mathrel{\spose{\lower 3pt\hbox{$\mathchar"218$}}
     \raise 2.0pt\hbox{$\mathchar"13C$}}}
\def\gta{\mathrel{\spose{\lower 3pt\hbox{$\mathchar"218$}}
     \raise 2.0pt\hbox{$\mathchar"13E$}}}
\def\Dt{\spose{\raise 1.5ex\hbox{\hskip3pt$\mathchar"201$}}}	
\def\dt{\spose{\raise 1.0ex\hbox{\hskip2pt$\mathchar"201$}}}	

\def\dotsfill{\leaders\hbox to 1em{\hss.\hss}\hfill}

\def\Gyr{{\rm\,Gyr}}

\slugcomment{Submitted to ApJ Letters}

\shorttitle{Ibata et al.}
\shortauthors{Halo white dwarfs}

\begin{document}

\title{Discovery of high proper motion ancient white dwarfs:\\
nearby massive compact halo objects?}


\author{Rodrigo Ibata}
\affil{Max-Plank Institut f\"ur Astronomie, 
K\"onigstuhl 17, D--69117 Heidelberg, Germany}
\email{ribata@mpia-hd.mpg.de}

\author{Michael Irwin}
\affil{Institute of Astronomy, Madingley Road, Cambridge, CB3 0HA, U.K.}
\email{mike@ast.cam.ac.uk}

\author{Olivier Bienaym\'e}
\affil{Observatoire de Strasbourg, 11 rue de l'Universit\'e,
Strasbourg, 67000 France.}
\email{bienayme@newb6.u-strasbg.fr}

\author{Ralf Scholz}
\affil{Astrophysikalisches Institut Potsdam, An der Sternwarte 16,
D--14482 Potsdam, Germany}
\email{rdscholz@aip.de}

\and

\author{Jean Guibert}
\affil{Centre d'Analyse des Images de l'INSU and Observatoire de Paris
(DASGAL/UMR-8633), 77 avenue Denfert-Rochereau, F-75014 Paris, France}
\email{Jean.Guibert@obspm.fr}



\begin{abstract}
We present the  discovery and spectroscopic identification  of two very high
proper motion ancient white dwarf stars, found in a systematic proper motion
survey. Their kinematics and  apparent magnitude clearly indicate that  they
are halo members,  while their optical spectra are  almost identical to  the
recently identified cool  Halo  white dwarf WD0346+246.  Canonical   stellar
halo models predict a white dwarf volume  density of two orders of magnitude
less than  the $\rho \sim  7\times 10^{-4} \msun\pc^{-3}$ inferred from this
survey.  With the caveat that the sample size is very small, it appears that
a significant fraction, $\sim 10$\%, of the local dark matter  halo is in the
form of very old, cool, white dwarfs.
\end{abstract}


\section{Introduction}

Unless our present understanding of gravitation is incorrect, the Milky Way
and most other galaxies, possess a halo of dark,  or barely luminous, matter
that extends well beyond  the  visible boundaries  of the  galaxy.   Present
cosmological constraints  strongly  favour non-baryonic dark  matter, though
the problem  is  complicated by the  possibility that  there  may be several
types of dark matter; for instance, galactic dark  matter may have little to
do with dark  matter in galaxy clusters or  on  larger scales \citet{bosma}.
In our own Galaxy, this Halo component appears to extend out  to at least 50
kpc as traced by the Magellanic Clouds \citep{lin}, and may well continue to
beyond  the orbit  of  Leo~I at  $\sim  200\kpc$ distance  \citep{zaritsky}.
Given the large  uncertainty in the extent of  these structures, their total
mass is  poorly constrained, but  even within 50  kpc it is clear  that they
must contain orders of magnitude  more mass than the clearly visible baryons
in the form of stars and gas.

One  of  the  few  direct  constraints  on  the  mass  distribution  of  the
constituent  particles of  galactic halos  comes from  the  Magellanic Cloud
microlensing experiments  \citep{alcock97, afonso}.  The  most recent report
of  the detection rate  of microlensing  events \citep{alcock00}  supports a
simple model  where approximately 20\% of the  Milky Way halo is  made up of
objects of  $\sim 0.5\msun$, consistent  with the result  of \citet{afonso}.
Though  the halo  fraction and  object mass  are model  dependent,  the main
uncertainty  arises  from  the  possibility  that  the  LMC  may  possess  a
substantial (dynamically hot) stellar halo of its own.  Given the absence of
evidence for  this latter possibility, the microlensing  conclusions must be
taken  seriously.  These  MACHOs  are  most  likely  situated  approximately
half-way to  the LMC,  where the optical  depth to microlensing  is highest.
However, unless  the MACHOs were effectively  decoupled from nucleosynthesis
(as would presumably be the case for primordial black holes), it is unlikely
that   MACHOs  fill   galactic   halos  entirely,   as   this  would   break
nucleosynthesis constraints.

Apart from ancient  white dwarfs (WDs), all stellar  MACHO candidates can be
safely  ruled out.   Ancient WDs,  which have  stellar masses  in  the range
measured by the MACHO experiment ($0.1$--$1.0 \msun$), are not yet ruled out
by direct starcount  observations, but are difficult to  envisage as a MACHO
population due to several indirect  constraints (see e.g., Fields, Freese \&
Graff  1998;  Graff  \etal\   1999),  including  most  importantly  chemical
enrichment  of  the  early  Galaxy  \citep{gibson}, though  for  a  possible
work-around of this problem, see \citet{chabrier}.  This renewed interest in
ancient  Halo WDs  is mainly  due  to a  theoretical reappraisal  of the  WD
cooling function.  New models  with a self-consistent treatment of radiation
transfer    and    the    inclusion    of   molecular    hydrogen    opacity
\citep{hansen,saumon}   drastically  change   the   predicted  colours   and
magnitudes of the  oldest, and hence coolest, WDs.   In particular, hydrogen
atmosphere (DA)  WDs with  ages $\simgt 10\Gyr$  have depressed red  and NIR
flux; contrary to previous expectations  they become bluer in, for instance,
${\rm V-I}$, with age.

\citet{ibata} recently  presented a proper motion analysis  using two epochs
of I-band  imaging data in the  Hubble Deep Field.  They  identified a small
sample 2--5 faint, and relatively  blue compact sources, that appear to move
with respect to the numerous background galaxies.  The colours, motions, and
number of  these sources  are consistent with  the hypothesis that  they are
members of a numerous halo population of ancient WDs, with a spectral energy
distribution similar to that predicted by the Hansen models.  The local mass
density of this population appears  similar to that of the ``standard'' halo
model   used  by   the  MACHO   and  EROS   collaborations  \citet{alcock97,
palanque-delabrouille}.  Unfortunately,  the measured proper  motions are at
the detection limit, and they are  exceedingly faint (${\rm V \sim 29}$), so
spectroscopic confirmation is currently impossible.

Striking confirmation that  the new WD models  are more appropriate has been
given  by \citet{hodgkin}.    Their  spectrum of WD0346+246   shows  a large
supression of the NIR flux relative to a black-body model fit to the visible
radiation, as predicted by \citet{hansen}  for ancient DA WDs.  Furthermore,
the proper motion of WD0346+246 and the  measured parallax, indicate that it
is most likely a Halo member.

Therefore,  Halo  WDs do  seem to  exist and  the DA  subset  have bluer
colours  than originally  expected (WD0346+246  has ${\rm  V-I=1.52}$).  The
next  important  question to  address  is the  local  mass  density of  this
population,  in  particular,  are  they  present in  sufficient  numbers  to
contribute sgnificantly to the proposed Halo of MACHOs?

\section{Survey}

To detect the nearby counterparts of the population of WD MACHOs tentatively
detected  by  \citet{ibata}, we  have  undertaken  complementary wide  field
photographic plate  surveys of very high  proper motion ($>1\scnp/\yr=$VHPM)
stars.   With a  survey limit  $\sim 10$  magnitudes brighter  than  the HDF
limit, we need to cover a field one million times larger than the $1.4\times
10^{-3} \Box^\circ$ WFPC2 field to  probe the same Halo volume. The expected
proper  motions   can  be  several  $\scnp/\yr$,  and   candidates  will  be
predominantly at  the faint limit  of the plates, where  existing catalogues
(cf.   LHS catalog,  Luyten 1979)  are heavily  incomplete.  Faced  with the
difficulty of reliably identifying  VHPM stars among thousands of potential
candidates, we have investigated three independent techniques to cross-check
results.   These analyses are  based on  different combinations  of existing
plate material and  probe different aspects of the  problem; survey 1:- uses
selected pairs of appropriate UK Schmidt Telescope (UKST) ${\rm B_J}$ plates
with short epoch differences for  sensitivity to ``blue'' VHPM stars; survey
2:- uses standard UKST ${\rm B_J}$ and  R sky survey plate pairs and has the
potential to probe all of the southern sky; and survey 3:- includes an extra
third epoch ESO-SRC  survey R plate to enable searching  to deeper limits
for $\delta < -17^\circ$.

All of  the southern sky has been  observed by the UKST  in both blue (${\rm
B_J}$) and red  (R) passbands, and the  designated survey quality plates for
more than half  of  this area have  been  processed by the  Automatic  Plate
Measuring  (APM) facility  \citep{kibblewhite}.   This  forms the basis  for
pilot studies using the  three survey methods.  The total error in measuring
the relative proper motion is generally $<<  0.1\scnp/\yr$, negligible in the
context of the survey.

\subsection{Survey 1}

While the generic  UKST  ${\rm B_J}$,   R survey  material  is  an excellent
resource for  locating objects   of average  colour  (g--k type)  and modest
proper  motions ($\approx0.5-2.0\scnp/\yr$),  it  may not  be ideal for  the
detection of ancient Halo WDs given the colour uncertainty and the generally
large   epoch difference, 10--15~years.  Consequently  for  survey~1 we have
based  proper  motion detection  on only  blue passband  (${\rm  B_J}$) UKST
plates with a shorter  epoch difference 1$<$ T  $<$ 10 years, to enhance the
sensitivity to the  oldest WDs.  As  would be  expected, the available  UKST
archival non--survey  grade  plates  are  a heterogeneous   mixture of plate
qualities  and epochs.  With the help  of Sue Tritton  of the  UKST Unit, we
selected   21 deep  ``b''-grade  ${\rm  B_J}$   southern  survey  plates for
scanning.  This  sample was chosen  in fields  with extant  APM on-line UKST
${\rm B_J}$ and OR catalogue data.

The  search  algorithm works as  follows:  first  the overall plate-to-plate
transformation is  derived using   all  objects  on the plates   that  match
(iteratively) within $2\scnp$.  The extra ``b''-grade ${\rm B_J}$ plate data
is then matched  to the exisiting survey ${\rm  B_J}$,R  catalogue.  Objects
matching within a   radius of $2\scnp$ for   our purposes are taken to  have
negligible proper motion and  are discarded.  This leaves  a catalog of what
are mostly spurious  detections (noise, diffraction spikes, extended  object
deblending problems,   close object   deblending   problems,  incompleteness
problems, etc.),  among  which there  is a  tiny  proportion of  actual VHPM
objects.   To   minimise  contamination by   spurious detections,  unmatched
objects were only considered as possible HPM objects if they were stellar in
appearance, and  brighter   than  ${\rm B_J}  =  22$  on  both  blue plates.
Remaining unmatched objects were  then considered to  be possible matches if
they were within a search radius of $30\scnp$.

In this way we obtained a provisional list of 1426 candidates.  After visual
examination of the  APM object finding  charts, DSS images  of the field (if
available), and visual comparision with the  R-band survey plates, we made a
sub-selection  of 101 candidate VHPM  stars for  followup observations.  The
majority of  the 1325 rejected  candidates were from situations  where there
was a clear error, for example, a detection on the APM  plate scan which did
not appear on the DSS scan of the same plate, detections near large galaxies
or bright stars.  We note that some of the  rejects at this stage were based
on less secure criteria having, for example, suspiciously elongated profiles
on the APM finding charts, or apparently different magnitudes after allowing
for the different plate depth between epochs.  Unfortunately, the probity of
rejection based    on   these last  criteria  is   difficult   to  quantify,
particularly near the plate limits. Two plates proved to be unusable, giving
a total area of 19 fields in this survey.

\subsection{Survey 2}

This difficulty in probing to near  the plate limits lead to our decision to
undertake the 2nd  complementary survey to shallower depths,  using just the
already available  APM online catalogue ${\rm  B_J}$ and R  plate data.  The
results from the  first search technique plus the properties  of the Halo WD
WD0346+246  were used to  fine tune  the selection  criteria.  In  a similar
manner to survey~1, brighter candidates,  stellar in appearance, with 12 $<$
R $<$ 19.5 and ${\rm B_J} <$ 21.5, and with possible PM $\simgt 1\scnp/\yr$,
were  selected from a  random sample  of 24  equatorial region  fields.  The
bright  limit was  imposed because  on deep  UKST sky  survey  plates images
brighter than  this are heavily  saturated, show strong  diffraction spikes,
and are imbedded in large  reflection halos, making both accurate photometry
and astrometry difficult to  achieve.  Equatorial fields were chosen because
on average they have much shorter epoch differences, $\sim 5$ years, between
the  R and  ${\rm B_J}$  survey plates  compared to  the  so-called Southern
survey plates, $\delta \le -20^\circ$, where the average epoch difference is
14 years.

Due to storage limitations at  the time of catalogue construction the extant
APM online catalogue recorded data on the coordinate system of the reference
R  plate and  only recorded  separate coordinate  information for  the ${\rm
B_J}$ plate  data if there are  no matches within $5\scnp$,  giving a lower
limit  to the  proper motions  that  can be  observed.  Of  the 24  selected
equatorial fields, 11 had epoch differences of 3 years or less, and although
processed were  not used in the  sample.  The remaining 13  fields had epoch
differences between 4--9 \yr\ and  produced a total of 87 candidates.  After
careful visual  inspection of the  online catalogue data 36  candidates were
left.  Finally, the remaining  candidates were cross-checked with DSS images
and visually inspected  on first epoch (1950s) Palomar  Sky Survey plates to
check the reality of the candidate and the proper motion.  This left a total
of 9 guaranteed VHPM candidates.

\subsection{Survey 3}

The third  strategy is based on a  three epoch plate analysis  from the UKST
APM on-line catalogue  and ESO-R survey plates scanned  by the MAMA (Guibert
et  al. 1984,  see also  http://dsmama.obspm.fr)  and is  more sensitive  to
bright, red, extreme HPM objects.  The survey covers 24 Schmidt plate fields
at  high  Galactic latitude,  selected  according  to  the high-grade  image
quality of plates and with epoch  differences between 3--15 \yr. Of these, 4
fields  are  in common  with  survey~1,  so  the effective  additional  area
surveyed is 20 fields.

Unlike the APM candidate selection which used the $x$--$y$ pixel coordinates
of the  scanned plates,  the ESO-R plate  MAMA data matching  used celestial
coordinates based on Tycho catalogue astrometric solutions \citep{robichon}.
VHPM  candidates  were selected  by  requiring  a  consistent alignement  of
celestial coordinates at the three  epochs.  The candidates were selected to
have  similar R magnitudes  on UKST-R  and ESO-R  plates.  Two  regimes were
considered: for ${\rm R < 18}$, PMs of $\mu <10\scnp/\yr$ were sought, while
for ${\rm  R > 18}$ a  PM upper limit  of $\mu < 1.5\scnp/\yr$  was imposed.
The 417  candidates thus  detected were inspected  visually on DSS-I  and II
images (when  available) and also ESO-R  plates, from which  20 objects were
selected  for observation,  corresponding  to the  bluest detections  (${\rm
B_J-R} < 2$) with the highest PM.

\section{Observations}

All  of the 101 candidate VHPM  stars detected in  survey~1 were imaged with
EFOSC2  on the ESO~3.6m telescope   on the night of   1999 October 1.  Short
g-band exposures (to match the $\rm B_J$ passband) were obtained centered on
the discovery positions.  Of the 101 candidates; 72 detections were found to
be  due  to  incompleteness  of the photographic    plates; 20 objects  were
confirmed  to  be  genuine VHPM stars,  while   a further 9  candidates were
consistent with objects that had moved out of the $5'$ field  of view of the
instrument  (that is, possibly  amazingly high PM  objects,  but most likely
just due to noise).

Of the 20 candidates  from survey~3, direct  CCD exposures revealed  that 18
were real moving  stars with $\mu<1\scnp/\yr$,  while the 2 most  extreme PM
candidates were rejected  being  in fact   due to spurious  alignments of  3
different   stars, each  one  appearing  alone  on each  plate,  an expected
occurence close to the detection limit of plates.

On the nights of 1999 October  2 and 3, four of  the 20 confirmed candidates
from survey~1 were observed with the spectroscopic mode of EFOSC2. Very high
winds limited our  choice of targets,  while the poor seeing conditions made
us choose to  set  the  slit  width  to   $2\scnp$, reducing  the   spectral
resolution.  All objects  were first observed  with grating \#1 (3185\AA\ to
10940\AA),  to obtain an  identification  spectrum, with follow-up with  the
higher resolution gratings \#9 (4700\AA\  to 6770\AA), and \#12 (6015\AA\ to
10320\AA).

Finally,  the  spectrum of  the  most extreme  high  proper  motion star  of
survey~2,  in   UKST  field  f821,  was   kindly  observed  for   us  by  L.
Storrie-Lombardi \& C. Peroux at the CTIO 4m on the night of 1999 October 13
using the R-C spectrograph with the Loral 3k CCD and the KPGL-2 grating covering
the wavelength range 3350\AA\ to 9400\AA, sampled at 2\AA /pixel.

\section{Results}

The  full  results of  these  surveys will  be  presented  in a  forthcoming
paper.  Here we focus on two VHPM stars discovered in surveys~1 and 2.

In the left panel of Figure~1, we display the flux-calibrated low resolution
spectrum  of  the  highest  proper  motion  star  in  survey~1,  F351-50,  a
featureless WD with ${\rm B_J=19.76}$, ${\rm B_J-R=1.65}$, ${\rm R-I= 0.31}$
(I-band photometry from DENIS)  and $\mu=2.33\scnp/\yr$ at PA $130.1^\circ$,
located  at $\alpha_{\rm  J2000}=0$  45 17.78,  $\delta_{\rm J2000}=-33$  29
10.0, epoch 1987.88 .  The upper panel of Figure~2 acts as a finder and also
demonstrates the high proper motion  of the object.  The absence of distinct
absorption lines  and the colour of  F351-50 are consistent with  it being a
cool  DA  WD.  (Unfortunately,  this  absence  of  spectral lines  makes  it
impossible  to measure  a radial  velocity for  this object,  even  from the
higher  spectral  resolution data).   Also  superimposed  in  Figure~1 is  a
$3500$~K black body model, which provides a good fit to the visual region of
the spectrum with $\lambda < 6500\Aa$.  It is readily apparent that there is
a substantial  depression of the flux  redward of $6500\Aa$  in this object,
precisely as was  originally seen in the now confirmed  ancient halo WD star
WD0346+246 (Hambly \etal\ 1997, their Figure~3).

What Galactic population does this object belong to?  Lacking a parallax, we
may still estimate its distance  from other constraints. First, the Universe
is not  old enough for $0.5\msun$ DA  WDs to have cooled  fainter than ${\rm
M_V = 18}$ \citep{hansen}  so F351-50 must be located at a  distance $d > 15
\pc$.  At this distance its total  space velocity with respect to the Sun is
$v > 170\kms$, which  clearly indicates that it is not a  thin or thick disk
member.   The  only remaining  alternative  is that  it  is  a Halo  object.
Assuming an  upper limit to the  space motion of $300\kms$,  its distance is
constrained to be  closer than $26.5\pc$ (this neglects  the radial velocity
component).   Finally, assuming, for  now, a  similar absolute  magnitude to
WD0346+246 places  it at a distance  of $\sim 25\pc$.  At  that distance the
expected   reflex   solar  motion   for   a   stationary   Halo  object   is
$\mu=1.95\scnp/\yr$ at PA $145.2^\circ$, in good agreement with the observed
motion.

Given the success  of this initial ${\rm B_J}$ plate  survey and the similar
colour  to  WD0346+246,  we  subsequently  checked whether  the  result  was
reproduced  in survey~2,  which  used  the extant  APM  catalog scans.   The
spectrum  of F821-07, the  most extreme  HPM star  found in  survey~2 (${\rm
B_J=18.91}$,  ${\rm B_J-R=1.16}$,  $\mu=1.72\scnp/\yr$ at  PA $201.2^\circ$,
$\alpha_{\rm  J2000}=23$ 19  09.96, $\delta_{\rm  J2000}=-6$ 12  32.5, epoch
1988.92), is shown in the right hand panel of Figure~1 together with finding
charts and demonstrable  high proper motion in the  lower panel of Figure~2.
F821-07  also has a  similar spectral  shape to  WD0346+246 \citep{hodgkin},
showing it  to be  another featureless  WD star, and  the very  large proper
motion, $\mu=1.72\scnp/\yr$ implies Halo  membership for the same reasons as
stated previously for F351-50.

Interestingly, after careful inspection of  the  LHS catalogue it was  clear
that F821-07  had the same proper motion  within measurement errors, and was
within 1~arcmin of the position of, the previously known HPM object LHS~542.
F351-50 has no plausible counterpart in the LHS catalogue  and remains a new
discovery.  As we will show in a forthcoming  paper, the population of known
nearby white dwarfs contains a previously unnoticed ancient Halo population,
and LHS~542,  in particular, is  an excellent candidate for  a cool Halo WD.
Furthermore, the known parallax of  LHS~542 implies a  distance of $31 \pm 3
\pc$ \citep{leggett} and hence a high space motion typical of Halo objects.

\section{Conclusions}

No ancient  WDs were found  in survey  3.  This is  partly due to  the added
incompleteness  from  the  requirement  that  a star  be  detected,  and  be
classified as stellar,  on 3, rather than 2, plates.  At  each epoch, for $R
\sim 19$, the  probability of detection and correct  classification is $\sim
85$\%. For  rapidly moving sources, there  is also a $\sim  5$\% chance that
they  will randomly  be situated  close (within  $3\scnp$) to  the detection
isophote of a  stationary source, and so not be detected  as a point source.
We  estimate the  completeness of  surveys~1 and  2 is,  respectively, $\sim
65$\% and $\sim 80$\%, while  survey~3 is $\sim 50$\% complete.  Survey~3 is
also  not sensitive to  faint ${\rm  R>18}$, high  PM $\mu>1.5\scnp/\yr$
stars: one of  the two detected HPM stars is outside  of the selection limit
of this survey.  This implies that  the efficiency of the detection is close
to  half  the   efficiency  of  the  first  two   strategies,  reducing  the
completeness of survey~3 to $\sim 25$\%.

Discounting the overlapping regions between   adjacent plates, and the  area
occupied by bright stars  and unusable regions of the  plates, surveys  1, 2
and 3 have, respectively, areas of $551\Box^{\circ}$, $377\Box^{\circ}$, and
$506\Box^{\circ}$,    giving a  total   effective    area  of  $\Omega  \sim
790\Box^{\circ}$ explored.

We  have only  followed up  obvious detections  up to  ${\rm  R=19}$.  Model
predictions (B. Hansen, priv.  comm.)   for $T>3500$~K DA WDs are that ${\rm
M_R<16.4}$, so we are sensitive  to a distance modulus of ${\rm m-M_0=2.7}$,
that is, a distance of $d=33\pc$.

The finding of $n=2$ stars closer than $d=33\pc$,  over the $\Omega \sim 790
\Box^\circ$ area of the survey implies a local density of:
\begin{equation}
\rho \sim n {{3} \over {\Omega d^{3}}}  M_\star \,
\end{equation}
where $M_\star$ are the individual object masses. Lacking any information on
individual  masses, we assume  a reasonable  value  of $M_\star \sim 1\msun$
\citep{chabrier}.  Thus the local mass density of this population is:
\begin{equation}
\rho \sim 0.0007 \msun\pc^{-3} \, .
\end{equation}
The expected mass density of white dwarfs in the stellar spheroid, given the
subdwarf  star   counts,  and  assuming   a  standard  IMF,   is  $1.3\times
10^{-5}\msun \pc^{-3}$  \citep{gould}, that is, approximately  two orders of
magnitude  lower than  this estimate.   This suggests  that the  IMF  is not
universal,  and was substantially  steeper in  the progenitor  population of
these WDs.

The  density derived  above is  $\sim 10$\%  of the  local  density ($0.0079
\msun\pc^{-3}$) of the ``standard'' dark matter halo model used by the MACHO
collaboration \citep{alcock97}.  However, the  halo mass fraction in WDs could
be even larger, as  surveys 2 and 3 are probably not  sensitive to the older
(and bluer) DA  WDs; while similar aged helium atmosphere  (DB) WDs would be
too faint to be detected by any  of these methods (see Hansen 1998). We note
that our  density estimate is consistent  with the updated  results from the
MACHO collaboration \citep{alcock00}, which appeared after the submission of
this paper.

Thus it appears  that a significant fraction of galactic  dark matter may be
baryonic, and in the form of cool and hence very old WDs, the remants of the
first stellar populations.   We caution that this analysis  is based on very
low  number statistics,  and  therefore requires  confirmation.  An  all-sky
survey to ${\rm R \sim 19}$ at Galactic latitudes $b > |30|$ (as is feasable
to undertake with extant photographic plates), and with completeness similar
to the  present work, should find  $\sim 14$ such stars,  providing a sample
with which the Halo may be  age-dated in a way independent of isochrone fits
\citep{richer}.  Also, very deep HST proper motion data towards the Galactic
center and anticenter directions  will constrain the radial density gradient
of  this population,  thereby  checking  whether this  is  a local,  perhaps
transient, feature, or a massive Galactic structure.

\begin{figure}
\includegraphics[width=\hsize]{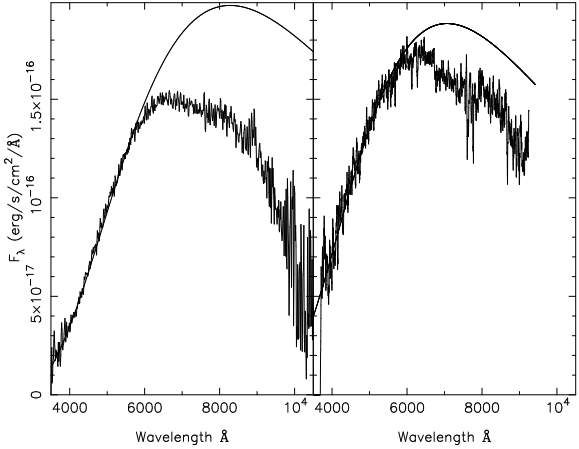}
\figcaption{The  flux-calibrated  spectra  of  the  two  VHPM  stars  (left:
F351-50,  right: F821-07).  The  black body  SED models,  fit to  $\lambda <
6500\Aa$) have  been overlaid.  The  black body temperatures are  3500~K and
4100~K for, respectively, F351-50 and F821-07.}
\end{figure}

\begin{figure}
\includegraphics[width=\hsize]{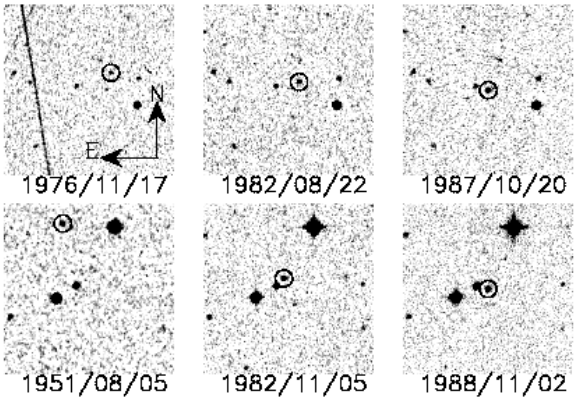}

\figcaption{Plate  direct  images   ($2\mcnd5\times2\mcnd5$)  of  (top  row)
F351-50 (from  left to right: ESO B,  UKST ${\rm B_J}$, UKST  R) and (bottom
row)  F821-07 (from  left to  right: Palomar  O, UKST  ${\rm B_J}$,  UKST R)
showing  proper motion.}
\end{figure}

\acknowledgements{We  would like  to thank  Sue Tritton  at UKSTU,  the MAMA
team, and our observers LSL and CP  at CTIO for their invaluable help. RI is
grateful  to  Strasbourg  Observatory  for  their  kind  hospitality  during
visits.}

\end{document}